\documentstyle[12pt,epsf]{article}
\newcommand{\be}{\begin{equation}}
\newcommand{\ee}{\end{equation}}
\newcommand{\ba}{\begin{eqnarray}}
\newcommand{\ea}{\end{eqnarray}}
\newcommand{\bb}{\begin{thebibliography}}
\newcommand{\eb}{\end{thebibliography}}
\newcommand{\ci}[1]{\cite{#1}}
\newcommand{\bi}[1]{\bibitem{#1}}
\newcommand{\lab}[1]{\label{#1}}
\begin{document}

\thispagestyle{empty}
\phantom{.}
\vspace{5mm}
\begin{flushright}
IFUP-TH  \hspace{2mm} 66/95
\end{flushright}
\vspace{3mm}

\begin{center}
{\bf STRUCTURE OF POMERON COUPLINGS AND SINGLE-SPIN ASYMMETRY IN
DIFFRACTIVE $Q \bar Q$ PRODUCTION
}\\[20mm]
{\large S.V.Goloskokov}$^{1} $\\
Dipartimento di Fisica dell'Universit\`a di Pisa\\
Instituto Nazionale di Fisica Nucleare, Sezione di Pisa\\
and Bogoliubov Laboratory of Theoretical  Physics,\\
  Joint Institute for Nuclear Research,
Dubna 141980, Moscow region, Russia.$^{2} $\\[20mm]
\end{center}

\begin{abstract}
Transverse
single spin asymmetry in polarized diffractive
$Q \bar Q$ production is calculated. It is shown that this asymmetry
depends strongly on the spin structure
of the pomeron coupling, which permits one to study the pomeron couplings
properties in future polarized experiments.
\end{abstract}

\vspace{20mm}
\begin{flushleft}
------------\\
$^{1} $ E-mail:  goloskkv@thsun1.jinr.dubna.su \\

$^{2} $ Permanent address

\end{flushleft}
\newpage

It is known that in high-energy reactions with polarized particles large spin
asymmetries are observed. A lots of information on the spin structure of
QCD can be obtained from double spin asymmetries
\ci{soff};
however, it is necessary to have two polarized particle beams (or a polarized
beam and a target) to study such asymmetries.
For theoretical and experimental investigations of spin effects in QCD, the
 single spin asymmetry should be very convenient. This problem
is very important due to extensive spin programs proposed at HERA, RHIC and
LHC (see e.g.\ci{now,bunce}).

Single spin asymmetry depends strongly on
the hadron properties. It is determined by the relation
\be
A=\frac{\sigma(^{\uparrow})-\sigma(^{\downarrow})}
{\sigma(^{\uparrow})+\sigma(^{\downarrow})}=
\frac{\Delta \sigma}{\sigma} \propto
 \frac{\Im (f_{+}^{*} f_{-})}{|f_{+}|^2 +|f_{-}|^2},  \lab{astr}
\ee
where $f_{+}$ and $f_{-}$ are spin-non-flip and spin-flip amplitudes,
respectively. So, single spin asymmetry appears if both $f_{+}$ and $f_{-}$
are nonzero and there is a phase shift between these amplitudes.
It can be shown that this asymmetry is of an order of magnitude
$$
A \propto \frac{m \alpha_s}{\sqrt{p_t^2}}.
$$
Here the mass $m$ can be about the
hadron mass \ci{ter}. So we can expect large transverse asymmetry for
$p_t^2 \simeq {\rm Few}~ GeV^2$. For such momenta transfer the diffractive
processes should be important.

The study of diffractive reactions has attracted considerable
interest due to the observation of high $p_t$ jets in diffractive
collisions \ci{gap}. Such processes where a proton is not broken
are determined at high energies by the pomeron
exchange. It can be interpreted as the
observation of partonic structure of a pomeron \ci{ing}.
These experiments have stimulated theoretical
investigations of diffractive reactions and pomeron properties.

The question about the spin
structure of the pomeron arises for the diffractive reactions with polarized
particles.
The experimental study of transverse spin asymmetries in these
processes shows that
at high energies and momentum transfer $|t| \ge 1 GeV^2$ they are not
small \ci{nur} and can possess a weak energy dependence.
This means that the pomeron can be complicated in the spin structure
\ci{zpc,akch}.

In this paper we shall calculate the transverse single-spin asymmetry in
high-energy diffractive $Q \bar Q$ production. It will be shown that this
asymmetry depends strongly on the spin structure
of pomeron couplings. The obtained effects permit one to study the spin
properties of the pomeron from the diffractive reactions in future
polarized experiments.

The pomeron is a colour--singlet vacuum $t$-channel exchange that can be
regarded
as a two-gluon state.
The pomeron contribution to the hadron high energy amplitude can be written
as a product of two pomeron vertices $V_{\mu}^{hhI\hspace{-1.1mm}P}$
multiplied by some function $I\hspace{-1.6mm}P$ of the pomeron.
As a result,  the quark-proton high-energy amplitude
looks as follows
\be
T(s,t)=i I\hspace{-1.6mm}P(s,t) V_{qqI\hspace{-1.1mm}P}^{\mu} \otimes
V^{ppI\hspace{-1.1mm}P}_{\mu}.    \lab{tpom}
\ee

In the nonperturbative two-gluon exchange model  \ci{la-na} and the
BFKL model \ci{bfkl} the pomeron couplings have a simple matrix
structure:
\be
V^{\mu}_{hh I\hspace{-1.1mm}P} =\beta_{hh I\hspace{-1.1mm}P} \gamma^{\mu},
\lab{pmu}
\ee
 which leads to spin-flip effects decreasing with energy as a power of $s$.
We shall call this form the standard coupling.

However, in some models the spin-flip effects do not vanish as
$s \to \infty$ \ci{zpc,models}.
It was shown that the pomeron--proton vertex is of the
form ( see \ci{zpc} e.g.)
\be
V_{ppI\hspace{-1.1mm}P}^{\mu}(p,r)=m p_{\mu} A(r)+ \gamma_{\mu} B(r),
\lab{prver}
\ee
where the amplitudes $A$ and $B$  are connected with the
proton wave function.
For the spin-average and longitudinal polarization of the proton
the term $B$ is predominant. So, the longitudinal double spin
asymmetry does not essentially depend on the pomeron-proton vertex structure.

The situation is drastically different for
spin asymmetries with a transversely polarized proton. In this case the
structure of the
pomeron-proton coupling  is significant and both the
functions, $A$ and $B$, contribute.
Really, in the model \ci{zpc} the  amplitudes  $A$ and $B$
have a phase shift. As a result, the single spin asymmetry
determined by the pomeron exchange
\be
A^h_{\perp} \simeq \frac{2m \sqrt{|t|} \Im (AB^{*})}{|B|^2} \lab{epol}
\ee
appears.
So, the knowledge of all spin
structures in the pomeron-proton vertex function is important here.
The model \ci{zpc} predicts that the asymmetry at $|t| \ge 1GeV^2$ can be
about $10 \div 15\%$.

The form of the quark-pomeron coupling has been studied in ref \ci{gol-pl}.
It was shown that in addition to the standard pomeron vertex
(\ref{pmu}) determined by the diagrams where gluons interact
with one quark in the hadron  \ci{la-na}, the large-distance gluon-loop
effects
should be important. These contributions are presented in Fig.1. They lead
to  new structures in the pomeron coupling that is similar,
e.g., to the anomalous magnetic momentum of a particle.
Really, if we calculate the gluon loop correction of Fig.1a for the
standard pomeron vertex (\ref{pmu}) and the massless quark,
 we obtain
\be
\gamma_{\alpha}(/ \hspace{-2.3mm} k+/ \hspace{-2.3mm} r) \gamma_{\mu}
/ \hspace{-2.3mm} k \gamma^{\alpha} \simeq -2[2(/ \hspace{-2.3mm} k+
\frac{/ \hspace{-2.3mm} r}{2}) k^\mu+i \epsilon^{\mu\alpha\beta\rho}
k_\alpha r_\beta \gamma_\rho \gamma_5],
\ee
where $k$ is a quark momentum, $r$ is a momentum transfer.
So, in addition to the $\gamma_\mu$ term, new structures immediately appear
from the loop diagram.
 The perturbative calculations \ci{gol-pl} of graphs, Fig.1,
give the following form for this vertex:
\be
V_{qqI\hspace{-1.1mm}P}^{\mu}(k,r)=\gamma_{\mu} u_0+2 m k_{\mu} u_1 +
2 k_{\mu}
/ \hspace{-2.3mm} k u_2 + i u_3 \epsilon^{\mu\alpha\beta\rho}
k_\alpha r_\beta \gamma_\rho \gamma_5+im u_4
\sigma^{\mu\alpha} r_\alpha.    \lab{ver}
\ee
The spin structure of the quark-pomeron coupling (\ref{ver}) is
drastically different from the standard one (\ref{pmu}).
Really, the terms
$u_1(r)-u_4(r)$ lead to the spin-flip in the quark-pomeron vertex in contrast
to the term $u_0(r) \gamma_\mu$.  The functions
$u_1(r) \div u_4(r)$ are proportional to $\alpha_s$. They are not small
at large $r^2$  \ci{gol4}. Note that
the phenomenological  vertex $V_{qqI\hspace{-1.1mm}P}^{\mu}$ with
$u_0$ and $u_1$ terms was proposed in \ci{klen}.

The new form of the pomeron--quark coupling (\ref{ver}) should modify various
spin
asymmetries in high--energy diffractive reactions \ci{klen,golasy}. It
has been
found from the analysis of longitudinal double spin asymmetries  \ci{golasy}
that the main contribution is determined by the terms
$u_0$ and $u_3$ in (\ref{ver}). The axial-like term
$V^{\mu}(k,r) \propto  u_3(r) \epsilon^{\mu\alpha\beta\rho}
k_\alpha r_\beta \gamma_\rho \gamma_5$ is  proportional to the momentum
transfer $r$  and it survives only in diffractive reactions where the pomeron
has a nonzero momentum transfer ($r^2=|t|$). So, this new $u_3(r)$ term
does not change the standard pomeron contribution to the proton structure
functions because here the pomeron momentum transfer is equal to zero.

Let us investigate the single transverse spin asymmetry in the
$p\uparrow p \to p+Q \bar Q+X$ reaction.
The standard kinematical variables look as follows
\be
s=(p_i+p)^2,\; t=r^2=(p-p')^2,\;x_p=\frac{p_i(p-p')}{p_i p},
\ee
where $p_i$ and $p$ are the initial proton momenta, $r=p-p'$ is the proton
momentum transfer and $x_p$ is a fraction of the initial proton momemtum
carried off by the pomeron. To be sure that the pomeron gives a contribution,
$x_p$ must be rather small.
In the case when all the energy of the pomeron goes into $Q \bar Q$
production \ci{lanj} the process is determined
by Fig.2. Here the planar diagrams where the pomeron
couples with one quark in the loop are shown. There are nonplanar
graphs in which gluons from the pomeron interact with different quarks
in the loop. Such effects, as a rule, do not exceed $10$ percent as compared
to the planar--diagram contributions (see e.g. \ci{goljp}).

In what follows we shall calculate the diffractive $ Q \bar Q$ production
using  the form (\ref{ver}) for the pomeron coupling.
It leads to higher nonphysical powers $(k_{\perp}^2)^N$ in traces
 where $k_{\perp}$
is a transverse part of the quark momentum in the loop. Such contributions
should be cancelled with the nonplanar diagrams that restore the gauge
invariance of the total amplitude. However, in the nonplanar propagators
the large
scalar products $W^2 \sim (kr) \propto x_p s$  should appeared that are
determined by the longitudinal component of the pomeron momentum.
As a result, we find:
$$ \frac{k_{\perp}^2}{k_{\perp}^2+M_Q^2} \to
\frac{k_{\perp}^2}{k_{\perp}^2+M_Q^2}-\frac{k_{\perp}^2}{k_{\perp}^2+W^2}=
\frac{k_{\perp}^2}{k_{\perp}^2+M_Q^2}\frac{1}{1+k_{\perp}^2/W^2};$$
where $M_Q$ is a quark mass.
Thus, the  nonplanar diagrams should modify the results
only for a large transverse
momentum $k_{\perp}^2 \sim W^2 \sim x_p s$.

So, the
expression  (\ref{ver}) can be regarded as an effective pomeron coupling.
The obtained results should be independent of the
nonplanar contribution at moderate momenta transfer
$k_{\perp}^2 \le 5 \div 10 GeV^2$.
The complete study of this
problem is extremely difficult and will be done later.

The cross sections $\sigma$ and $\Delta \sigma $ determined in (\ref{astr})
can be written in the form
\be
\frac{d \sigma(\Delta \sigma)}{dx_p dt dk_{\perp}^2}=\{1,A^h_{\perp}\}
\frac{\beta^4 |F_p(t)|^2 \alpha_s}{128 \pi s x_p^2}
\int_{4k_{\perp}^2/sx_p}^{1} \frac{dy g(y)}{\sqrt{1-4k_{\perp}^2/syx_p}}
\frac{ N^{\sigma(\Delta \sigma)}
(x_p,k_{\perp}^2,u_i,|t|)}{(k_{\perp}^2+M_Q^2)^2}. \lab{si}
\ee
Here $g$ is the gluon structure function
of the proton, $k_{\perp}$ is a transverse momentum of jets, $M_Q$
is a quark mass, $N^{\sigma(\Delta \sigma)}$ is a trace over the quark loop,
$\beta$ is a pomeron coupling constant, $F_p$ is a pomeron-proton form factor.
In (\ref{si}) the coefficient equal to unity appears in $\sigma$ and
the transverse hadron asymmetry $A^h_{\perp}$ in the pomeron-proton vertex
determined in (\ref{epol}) appears in $\Delta \sigma$. We calculate
the traces using the programme REDUCE and the integrals by the programme
MAPLE.

The main contributions to $N^\sigma(N^{\Delta \sigma})$  in the discussed
region are determined by $u_0$ and $u_3$ structures in (\ref{ver}).
They can be written for $x_p=0$ in the form:
\ba
N^{\Delta \sigma}=16(k_{\perp}^2+|t|) k_{\perp}^2 u_0^2+\Delta N^{\Delta
\sigma}; \nonumber\\
N^{\sigma}=32(k_{\perp}^2+|t|) k_{\perp}^2 u_0^2+\Delta N^{\sigma}.
\lab{nsy}
\ea
where
\ba
\Delta N^{\Delta\sigma}=8[(k_{\perp}^2+|t|)u_3-2u_0]
(k_{\perp}^2+|t|) k_{\perp}^2 |t| u_3; \nonumber \\
\Delta N^{\sigma}=16[(k_{\perp}^4+4k_{\perp}^2|t|+|t|^2)u_3-2
(2k_{\perp}^2+|t|) u_0] k_{\perp}^2 |t| u_3.\lab{a}
\ea
Note that $u_3<0$.

The $k_{\perp}^2$ dependence  of $u_i$ functions which is important in the
calculation has been studied. It was found that all functions decrease
with growing $k_{\perp}^2$. A good approximation of this behaviour is
\be
u_i(k_{\perp},r)= \frac{|t|}{k_{\perp}^2+|t|} u_i(0,r),\;\;\;r^2=|t|.
\ee
This improves the convergence of the integral  over $d^2 k_{\perp}$.
The simple form of the $u_0(r)$  function
$$   u_0(0,r)=\frac{\mu_0^2}{\mu_0^2+|t|} $$
was used, with $\mu_0 \sim 1Gev$ introduced in  \ci{don-la}. The perturbative
QCD
results \ci{gol4} for the functions
$u_1(0,r) \div u_4(0,r)$ at $|t|>1 GeV^2$ have been used.

Both $\sigma$ and $\Delta \sigma$ have a similar
behaviour at small $x_p$
$$ \sigma(\Delta \sigma) \propto \frac{1}{x_p^2} $$
This important property of (\ref{si}) allows one to study asymmetry at small
$x_p$ where the pomeron exchange is predominated because of a high energy
in the quark-pomeron system.

We shall calculate integrals (\ref{si}) using the simple form
for the gluon structure function
\be
 g(y)=\frac{R}{y} (1-y)^5,\;\;\;\;R=3.  \lab{glstr}
\ee
This form corresponds to the pomeron with $\alpha_{I\hspace{-1.1mm}P}(0)=1$.
Just the same approximation for the pomeron exchange has been used in
calculations. The analysis can be performed for the  pomeron with
$\alpha_{I\hspace{-1.1mm}P}(0)=1+ \delta $ ($\delta>0 $) and more complicated
gluon structure functions but it does not change the results drastically.

In the  diffractive jets production studied here the main contribution
is determined by the region where the quarks
in the loop are not far of the mass shell.
Then the interaction time should be long and the pomeron
rescatterings can be important. They change properties of single
pomeron exchange. This type of the pomeron is called usually the
"soft pomeron" \ci{softp}. It can possess a spin-flip part with the
phase different from the spin-non-flip amplitude \ci{zpc}.
So, we can assume that the hadron asymmetry factor in (\ref{si}) can be
determined by the soft pomeron
and it coincides with the elastic transverse hadron asymmetry
(\ref{epol}). In our further estimations we shall use the magnitude
$A^h_{\perp}=0.1$.

The calculations were performed for $\beta=2GeV^{-1}$
\ci{don-la}
and the exponential form of the proton form factor
$$ |F_p(t)|^2=e^{bt} \;\;\;{\rm with}\;\; b=5GeV^2.$$

Our predictions for $\sigma$ and single spin asymmetry
for the energy of the
future fixed--target polarized experiments using the proton beam at
HERA-(HERA-N) -$\sqrt{s}=40GeV$, $x_p=0.05$ and $|t|=1GeV^2$
for the standard  (\ref{pmu}) and spin-dependent quark-pomeron vertex
(\ref{ver}) are shown
in Figs.3, 4 for the light--quark jets. It is easy to see
that the shape of asymmetry is different for the standard and
spin-dependent pomeron vertex. In the first case it is approximately constant,
in the second it depends on $k^2_{\perp}$, due to
the additional $k^2_{\perp}$ terms which appear in
$\Delta N^{\Delta \sigma}$ and $\Delta N^{\sigma}$ in (\ref{nsy}) for the
pomeron coupling (\ref{ver}).

We calculate the  cross sections $\sigma$ and
$\Delta \sigma$ integrated over $k^2_{\perp}$ of jets, too.
The cross sections $\sigma$ and $\Delta \sigma$ can be written in the form
\be
\frac{d \sigma(\Delta \sigma)}{dx_p dt}=\{1,A^h_{\perp}\}
\frac{\beta^4 |F_p(t)|^2 \alpha_s}{128 \pi s x_p^2}\; S\;(\Delta S)
\ee
\ba
S = 2 S_0+S_1;      \nonumber   \\
\Delta S = S_0 +\Delta S_1,
\ea
where $S_0$ is a contribution of the standard pomeron vertex (\ref{pmu})
\be
S_0 \simeq 8 [2 \ln(\frac{H}{|t|}) + \ln(\frac{|t|}{M_Q^2})]
\ln(\frac{|t|}{M_Q^2}) R |t| u_0^2,        \lab{s0}
\ee
and $S_1, \Delta S_1$ are determined by $u_0$ and $u_3$ terms in the
spin-dependent pomeron coupling (\ref{ver})
\ba
S_1 \simeq \frac{8}{3}[6(\ln(\frac{|t|}{M_Q^2})+2) \ln(\frac{H}{|t|})
|t| u_3 -12 (\ln(\frac{|t|}{M_Q^2}) + 1) \ln(\frac{H}{|t|}) u_0 \nonumber \\
+3 (|t| u_3-2 u_0) \ln(\frac{|t|}{M_Q^2})^2 +3 \ln(\frac{H}{|t|})^2 |t| u_3
+ \pi^2 |t| u_3]
R |t|^2 u_3; \nonumber   \\
\Delta S_1 \simeq  \frac{S_1}{2}-16 ( |t| u_3-  u_0) \ln(\frac{H}{|t|})
 R |t|^2  u_3.    \lab{s1}
\ea
Here $H=sx_p/4$. In (\ref{s0},\ref{s1}) the parts of $S$ and $\Delta S$
determined by the residue $R$ in the gluon structure function (\ref{glstr})
are written. In the calculations, a more complicated form of $S$ and
$\Delta S$ has been used.

Our predictions for $\sigma$
 at the energy $\sqrt{s}=40GeV$, $x_p=0.05$ for the standard  vertex
 (\ref{pmu}) and  spin-dependent quark-pomeron coupling (\ref{ver}) are shown
in Fig.5 for light--quark jets. The obtained $|t|$ -dependence of the
cross-section is very similar in both the cases, but $\sigma$ is larger for
the vertex (\ref{ver}).

The asymmetry obtained from these integrated cross
sections does not practically depend on the quark-pomeron vertex structure.
It can be written in both the  cases  in the form
\be
A1=\frac{\int dk^2_{\perp} \Delta  \sigma}{\int dk^2_{ \perp}\sigma} \simeq
0.5 A^h_{\perp}  \lab{a1}
\ee
As a result, the integrated asymmetry (\ref{a1}) can be used for studying
the hadron asymmetry $A^h_{\perp}$ caused by the pomeron.

Thus, in this paper, the perturbative QCD analysis of single spin asymmetry
in diffractive 2-jet production  in the $pp$ reaction is performed
using the spin-dependent pomeron coupling.
The nonplanar graphs eliminated from our consideration  can change
the $k^2_{\perp}$
dependence of distributions at large $k^2_{\perp}$.
The obtained results should be true up to
$k^2_{\perp} \sim 5 \div 10 GeV^2$.
The estimated errors in the measured asymmetry \ci{golhera} show that the
structure of the quark-pomeron vertex can be studied from the $k^2_{\perp}$
distribution of single-spin asymmetry in this momentum--transfer region
at HERA-N \ci{now}.
Integrated cross sections are determined mainly by the region
$k^2_{\perp} \le |t|$. They do not practically depend  on the
contribution of nonplanar graphs and can be used to extract  information
on the pomeron-hadron coupling.

The asymmetries discussed here have a weak energy dependence. For the
energies of RHIC and LHC fixed--target experiments ($\sqrt{s}=120GeV$)
they are not far from our results at $\sqrt{s}=40GeV$. So, the diffractive
polarized experiments
at HERA-N, RHIC and LHC energies permit one to study spin
properties of quark-pomeron and proton-pomeron vertices determined
by QCD at large distances.
\\

I am grateful to A.V.Efremov, P.Kroll, W.-D.Nowak, G.Ramsey, A.Penzo,
A.Sch\"afer, O.V.Teryaev for fruitful
discussions.
Special thanks to A.Di Giacomo and M.Mintchev Dipartimento di Fisica
dell'Universit\`a di Pisa,
Instituto Nazionale di Fisica Nucleare, Sezione di Pisa,
 P.Colangelo and D.Nardulli
Instituto Nazionale di Fisica Nucleare, Sezione di Bari for critical remarks
and hospitality at Pisa and Bari where this paper has been
completed.

This work was supported in part by the Russian Foundation for
Fundamental Research, Grant 94-02-04616.
\\
%%%%%%%%%%%%%%%%%%%%%%%%%%%%%%%%%%%%%%%%%%%%%%%%%%%%%%%%%%%%%%%%%%%%%%%%%%
\newpage
%%%%%%%%%%%%%%%%%%%%%%%%%%%%%%%%%%%%%%%%%%%%%%%%%%%%%%%%%%%%%%%%%%%%%%%%%%

% \end{document}

\newpage

  \vspace*{2.3cm}
\epsfxsize=8cm
\centerline{\epsfbox{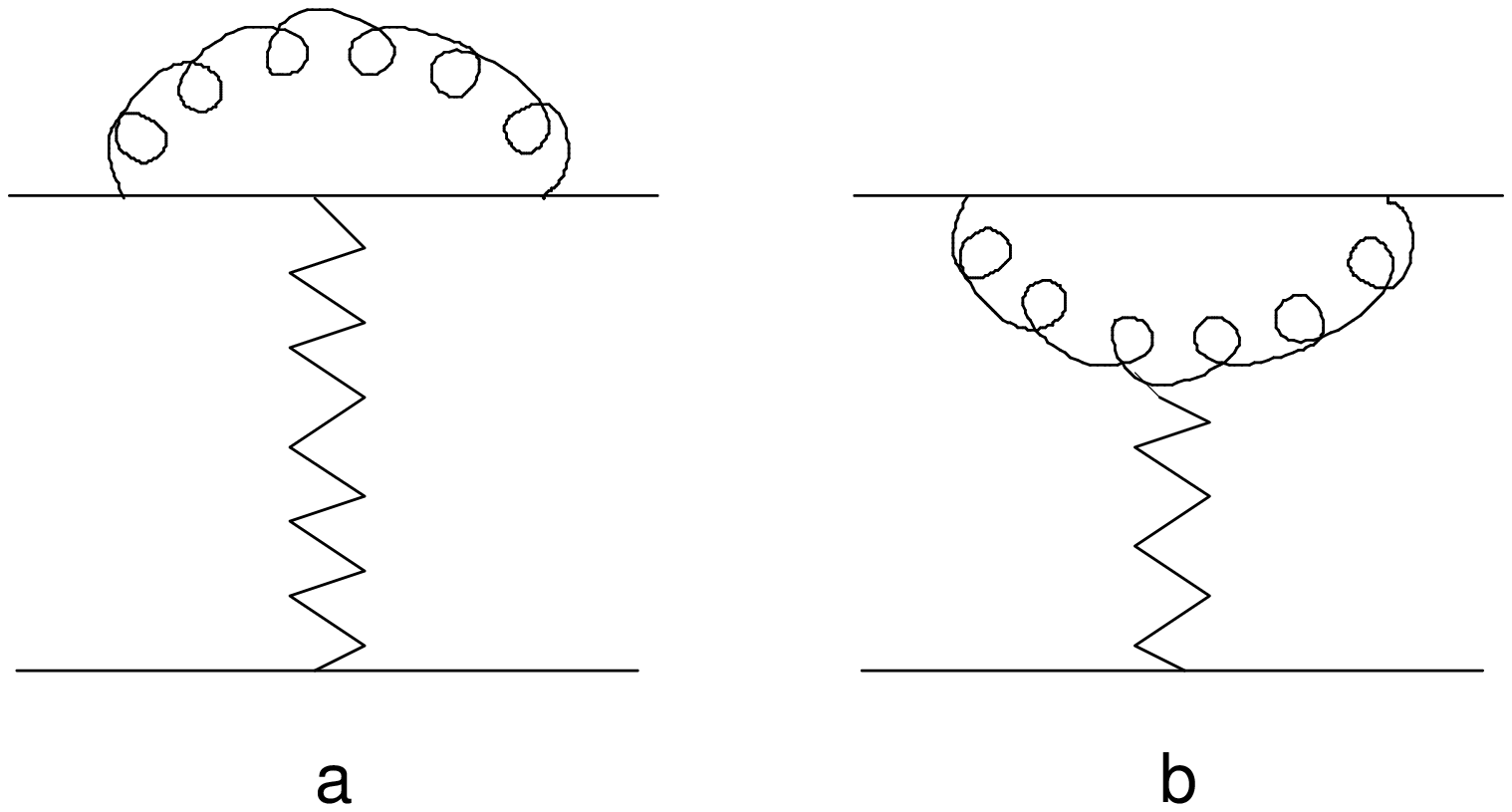}}
  \vspace*{-6.5cm}
Fig.1 ~Gluon-loop contribution to the quark-pomeron coupling.
Broken line -the pomeron exchange.

\samepage
% \begin{center}
  \vspace*{2.5cm}
      \hspace*{4.1cm}
\mbox{
   \epsfxsize=12cm
   \epsffile{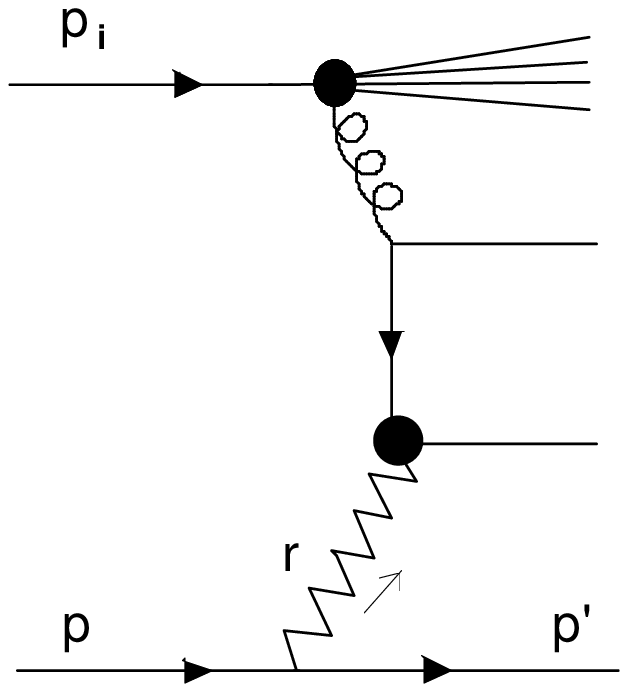}
}
%\end{center}
  \vspace*{-10.5cm}

Fig.2: ~Diffractive $Q \bar Q$ production in $pp$ reaction.

\newpage
  \vspace*{-.5cm}
\epsfxsize=12cm
\centerline{\epsfbox{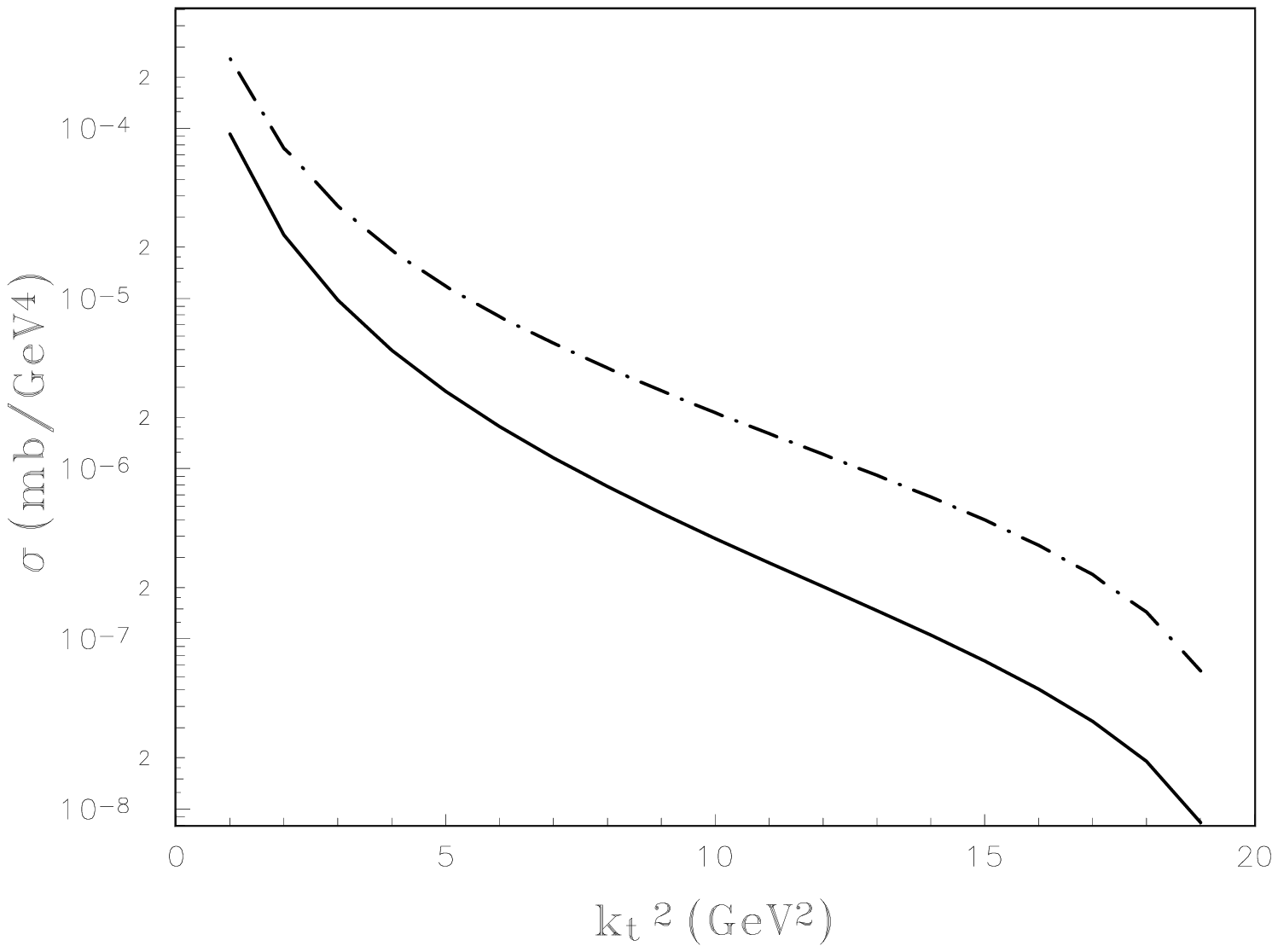}}
  \vspace*{-.2cm}
Fig.3 ~Distribution of $\sigma$ over jets $k^2_{\perp}$.
Solid line -for standard vertex;
dot-dashed line -for spin-dependent quark-pomeron vertex.

 \begin{center}
  \vspace*{-.2cm}
\mbox{
   \epsfxsize=12cm
   \epsffile{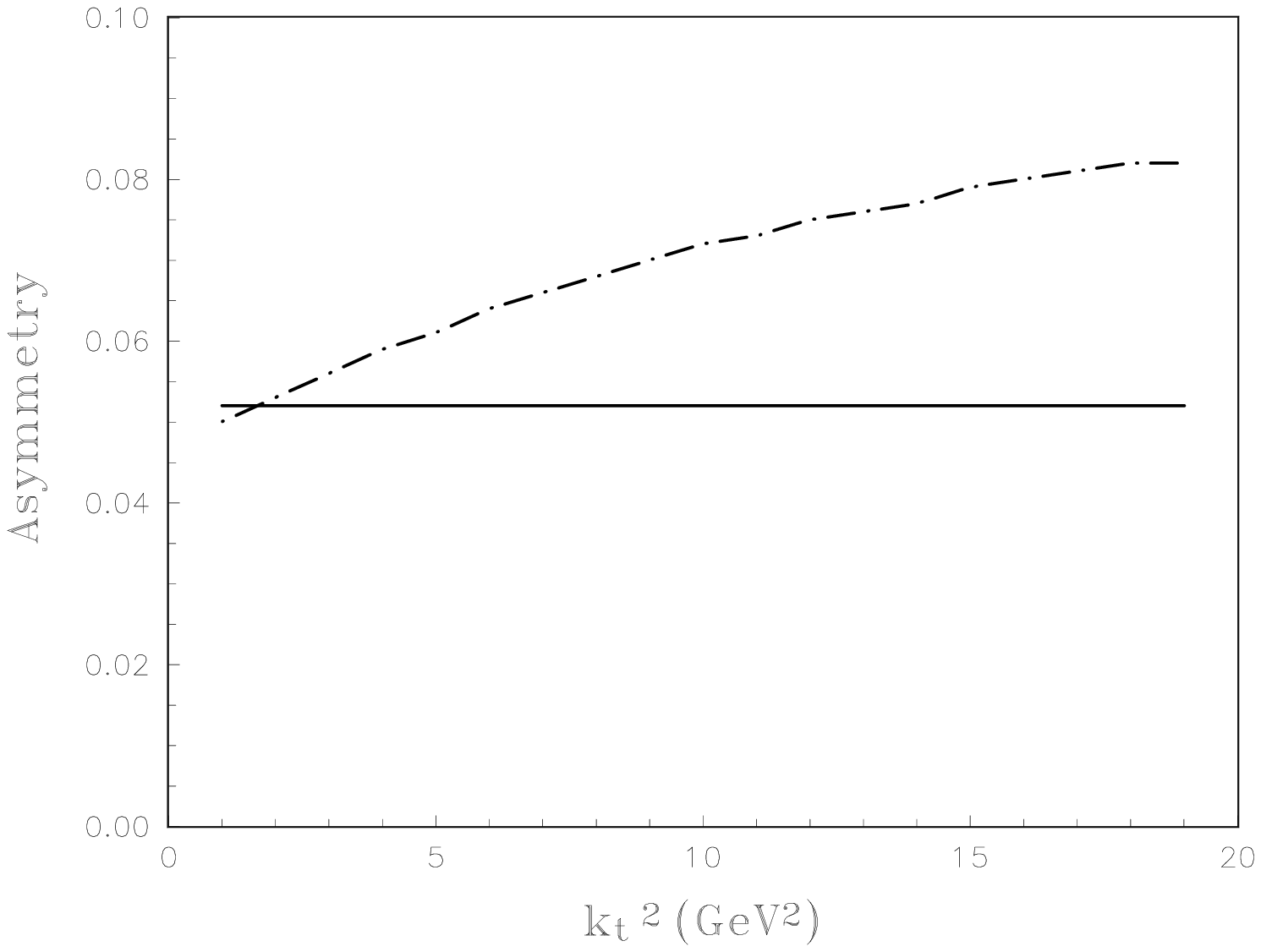}
}
\end{center}

  \vspace*{-.5cm}
Fig.4: ~Distribution of asymmetry over jets $k^2_{\perp}$.
Solid line -for standard vertex;
dot-dashed line -for spin-dependent quark-pomeron vertex.

\newpage
  \vspace*{-.5cm}
\epsfxsize=12cm
  \centerline{\epsfbox{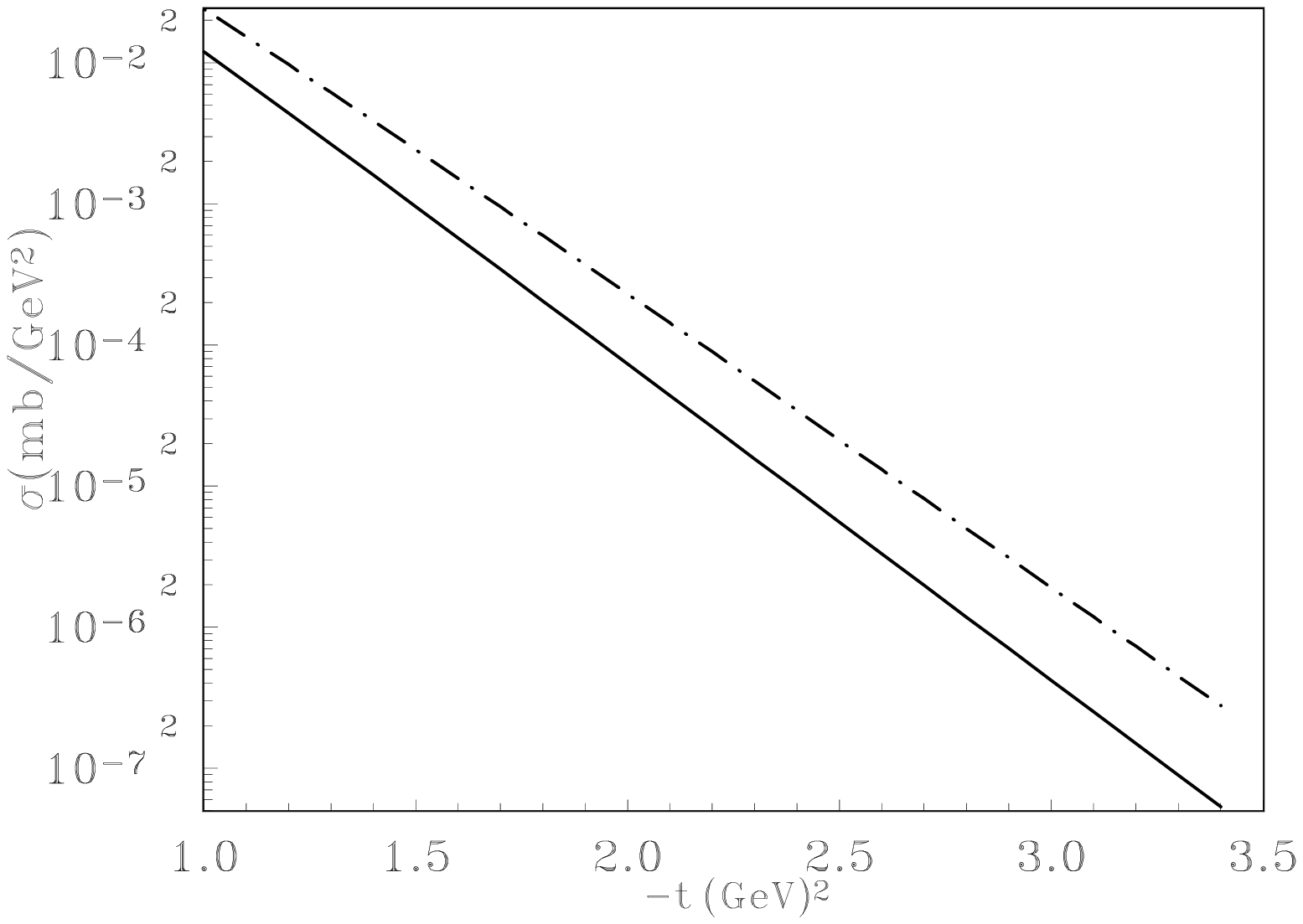}}
  \vspace*{-.2cm}
Fig.5 $|t|$-- dependence of cross section $\sigma$
integrated over jet $k^2_{\perp}$.
Solid line -for standard vertex;
dot-dashed line -for spin-dependent quark-pomeron vertex.

 \end{document}